\documentclass[12pt]{article}
\begin{document}
\input{epsf}
\Large
\begin{center}   
{\bf
      Study of the hadronization process in }
\end{center}
\begin{center}
\bf{
    cold nuclear medium}
\end{center}

\normalsize
\begin{center}
      N.~Akopov\footnote{ supported by ISTC Grant A-1606}, 
      L.~Grigoryan\footnote{ supported by DESY, Deutsches 
Elektronen Synchrotron}, Z.~Akopov
\end{center}

\begin{center}
      Yerevan Physics Institute, Br.Alikhanian 2, 375036 Yerevan, Armenia
\end{center}

\begin{center}
\end{center}

\begin {abstract}

\hspace*{1em}
The improved two-scale model is used to perform the fit to the 
semi-inclusive deep-inelastic scattering (SIDIS) data of HERMES
experiment at DESY on nuclear targets. The ratio of hadron multiplicity on
nuclear target to the deuterium one is chosen as observable, as
usually. The two-parameter's fit gives satisfactory agreement with the data in
term of $\chi^2$ criterium. Best 
values of parameters are then used to calculate the nuclear multiplicity ratio
for the hadrons not included in the fit procedure.
\end {abstract}
\twocolumn  
\section{\label{sec:int} Introduction}
\normalsize 
Hadronic reactions in a nuclear medium, either cold or hot can shed additional
light on the hadronization process. Numerous measurements of hadron
production on nuclear targets in SIDIS of leptons~\cite{osborne}-\cite{brooks}
are available. In ultra relativistic heavy-ion collisions the jet-quenching and
parton energy-loss phenomena are observed~\cite{adcox,j.adams}. In each case
the observed hadron yields are differed from those in the corresponding reactions
on free nucleons. In comparison with other reactions leptoproduction has the
virtue that energy and momentum of the struck parton are well determined, as
they are tagged by the scattered lepton. Study of hadron production in SIDIS
on nuclear targets offers an opportunuty to investigate the quark (string,
color dipole) propagation in nuclear matter and the space-time evolution of
the hadronization process. If the final hadron is formed inside the nucleus,
it can interact via the relevant hadronic cross section, causing further
reduction of the hadron yield. The perturbative QCD cannot describe
hadronization process because of the essential role of "soft" interactions.
Therefore, the understanding of this process on the phenomenological level is
of basic importance for development of the theory. For this purpose we
investigate the nuclear attenuation (NA), which is a ratio of differential
hadron multiplicity on a nucleus to that on deuterium.\\
\begin{eqnarray}
\nonumber
R_{M}^{h}(\nu,z) = \frac{\Big(\frac{N^{h}(\nu,z)}{N^{e}(\nu)} \Big)_{A}}
{\Big(\frac{N^{h}(\nu,z)}{N^{e}(\nu)} \Big)_{D}} \hspace{0.2cm},
\end{eqnarray}
where $z = E_h/\nu$, $E_h$ and $\nu$ are energies of the final hadron and
virtual photon respectively,
$N^{h}(\nu,z)$ is the number of semi-inclusive hadrons at given
$\nu$ and $z$ and $N^{e}(\nu)$ is the number of inclusive DIS leptons
at given $\nu$. Subscripts $A$ ($D$) denote that reaction takes place
on nucleus (deuterium) respectively.
In the above formula more variables like the photon virtuality - $Q^2$, transverse hadron momentum in
respect to the virtual photon direction - $p_t$, over
which the NA is averaged,  are not written.
At present,
several phenomenological models for description of the
NA~\cite{bialas1}-\cite{sassot} are available.
The simple version
of the string model, so called Two-Scale Model (TSM), was proposed
by European Muon Collaboration
for the description of its experimental data~\cite{ashman}. In
Ref.~\cite{akopov2} improved version of TSM (ITSM) was proposed.
In present work ITSM is used to perform a fit to the recent
SIDIS data of HERMES experiment
on nuclear targets~\cite{airapet4}. For a fit we use
the more precise (high statistic) part of the data sample including
one dimensional data for $\pi^+$ and $\pi^-$ mesons and two dimensional
data for charged pions.
One (two) dimensional data means that data are presented in form of
function of one (two) variable.
Then the $R_{M}^{h}$ for all measured hadrons
were calculated with the values of parameters corresponding to the minimum
values of reduced $\hat{\chi}^2=\chi^{2}/d.o.f.$ (here d.o.f. denotes "degree of freedom"). 
Then the results of such "best fit" were compared with the experimental data.\\
\hspace*{1em}
The remainder of the paper is
organized as follows. In section~\ref{sec:itsm} we briefly remind 
about the ITSM.
In section~\ref{sec:fit} the part of data included in fit and
some details of fitting procedure are presented.
Results are discussed and compared both with the different versions of
present fit
and with our preceding one~\cite{akopov2}.
In section~\ref{sec:res} we compare results of the fit with experimental
data and discuss them.
Conclusions are given in section~\ref{sec:conc}.
\section{\label{sec:itsm} ITSM}
\normalsize
\hspace*{1em}
Basic formula for NA in TSM~\cite{ashman} is:  
\begin{eqnarray}
\nonumber
{R_A = 2\pi\int_0^{\infty} bdb \int_{-\infty}^{\infty}dx \rho(b,x)\times}
\end{eqnarray}
\begin{eqnarray}  
{[1-\int_x^{\infty} dx' \sigma^{str}(\Delta x)\rho (b,x')]^{A-1}} \hspace{0.2cm},
\label{eq:equ1}
\end{eqnarray}
where $b$ is the impact parameter, $x$ - longitudinal coordinate of the
DIS point, $\rho(b,x)$ - nuclear density function, $x'$- longitudinal
coordinate of the string-nucleon interaction point,
$\sigma^{str}$($\Delta x$) - the string-nucleon cross section on distance 
$\Delta x = x'-x$ from DIS point, $A$ - atomic mass number.
The above equation does not take into account the final state
interactions (FSI) in deuterium. In this paper we use,
following~\cite{akopov2},
more precise formula for the ratio of
multiplicities $R_{M}^{h} = R_{A}/R_{D}$.
The string model is based on the idea that after DIS the knocked out 
(anti)quark does not leave the nucleon remnant, but forms a string 
(color dipole) with the
(anti)quark on the fast and the nucleon remnant on slow ends, while the
color string itself consists of gluons. Its longitudinal size must be
larger than the transverse one, but cannot be essentially larger than
the hadronic size because of confinement. The string can break down
into two strings according to the following scenarios. First, when the
quark-antiquark pair from the color field of the string is produced; and
second, when the color interaction between the string and the nucleon
(lying on its trajectory) has happened (see for
instance~\cite{gyulassy,akopov4}). In the
"history" of the string there are two time scales which are of interest
to us. They are the time scales connected with the production of the 
first constituent (anti)quark of the final hadron and interaction of its
two constituents for the first time. 
\begin{figure}
\begin{center}
\epsfxsize=7.cm
\epsfbox{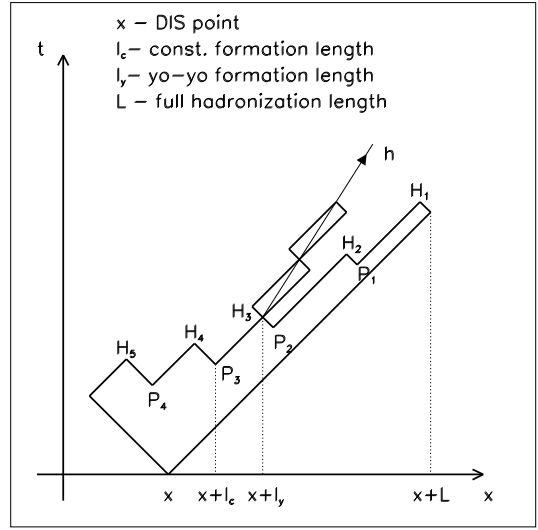}
\end{center}
\caption{\label{fig:fig1}  
{\it Space-time structure of hadronization in the string model. The two
constituents of the hadron are produced at different points. The
first and second constituents of hadron h are created at the points
P$_3$ and P$_2$, respectively.
They meet at H$_3$ to form the hadron.}}
\end{figure}
These two scales are (see
Fig.~\ref{fig:fig1}): $\tau_c$ ($l_c$) - constituent formation time
(length)\footnote{in relativistic units ($\hbar$ = c = 1, where $\hbar =
h/2\pi$ is the Plank reduced constant and c - speed of light) $\tau_i =
l_i$, i=c,h because partons and hadrons move with near light speeds.};
and $\tau_h$ ($l_h$) - yo-yo formation time (length). The yo-yo formation
means, that the  colorless system with valence content and quantum
numbers of the final hadron is formed, but without its "sea" partons.
In the two-dimensional string model which satisfy the following
conditions: (i) quark-antiquark pairs arising from the vacuum do not
have energy; (ii) energy loss of the leading quark on unit length
(string tension) is constant (widely known example is the Lund model),
there is a simple connection between $\tau_h$ and $\tau_c$
\begin{eqnarray}
\tau_h - \tau_c = z\nu/\kappa \hspace{0.2cm}, 
\label{eq:equ2}                    
\end{eqnarray}     
where
$\kappa$ - string tension (string
constant). Further we will use two different expressions for $\tau_c$. 
The first expression is obtained for hadrons containing
leading quark~\cite{kopel2}:
\begin{eqnarray}
\tau_c =(1 - z)\nu/\kappa  \hspace{0.2cm}.
\label{eq:equ3}                                   
\end{eqnarray}
The color string fully spends its energy on the distance of $L=\nu/\kappa$
beginning from the DIS point (see Fig.~\ref{fig:fig1}). Last hadron 
producing from the string is $h=H_1$, which contains leading quark and
carries energy $E_h$. At distance $L$, the energy of the leading quark
becomes equal to zero and whole energy of hadron is concentrated in
another constituent. This constituent collects its energy from the string,
and will have energy $E_h$ on distance $L$ only if it was produced on
distance $E_h/\kappa=z\nu/\kappa$ from $L$. This is reflected in
eq.~(\ref{eq:equ3}). It is important to note that the hadron produced 
on the fast end of string is not always necessarily the fastest hadron.
Second expression for $\tau_c$ used in this paper is its average value:
\begin{eqnarray}
\tau_c = \int_0^{\infty} ldlD_c(L,z,l)/\int_0^{\infty} dlD_c(L,z,l)
\hspace{0.2cm},
\label{eq:equ4}                                                       
\end{eqnarray}
where $D_c(L, z, l)$ is the distribution of the constituent formation
length $l$ of summed over all ranks hadrons carrying momentum $z$.
This distribution in framework of the standard Lund model~\cite{andersson}
was obtained in Refs.~\cite{czyzew,bialas2}:
\begin{eqnarray}
\nonumber
D_c(L,z,l) = L(1+C)\frac{l^C}{(l+zL)^{C+1}}\times
\end{eqnarray}
\begin{eqnarray}
\nonumber
\Big(\delta(l-L+zL)+\frac{1+C}{l+zL}\Big)\times
\end{eqnarray}  
\begin{eqnarray}
\theta(l)\theta(L-zL-l)  \hspace{0.2cm},
\label{eq:equ5}
\end{eqnarray}
where $C=0.3$ is the parameter which controls the
steepness of the standard Lund fragmentation function. The path traveled
by the string between the DIS and interaction points is $\Delta x = x'-x$.
In the TSM the string-nucleon cross section has form:
\begin{eqnarray}
\nonumber
\sigma^{str}(\Delta x) = \theta(\tau_c - \Delta x)\sigma_q + \theta(\tau_h -
\end{eqnarray}
\begin{eqnarray}
\Delta x)\theta(\Delta x - \tau_c)\sigma_s + \theta(\Delta x - \tau_h)\sigma_h
 \hspace{0.2cm},
\label{eq:equ6}
\end{eqnarray}  
where $\sigma_q$, $\sigma_s$ and $\sigma_h$ are the cross sections for 
interaction with the nucleon of the initial string, open string (the
string containing first constituent (anti)quark of final hadron
on its slow end) and final hadron, respectively.
In this model the string-nucleon cross section is a function 
which jumps in points $\Delta x = \tau_c$ and $\tau_h$.\\
\hspace*{1em}
In reality the string-nucleon   
cross section starts to smoothly increase from the DIS point, and reaches the
value of the hadron-nucleon one at $\Delta x$ = $\tau$.\\
\hspace*{1em}
Unfortunately, it is impossible to obtain $\sigma^{str}$
from perturbative QCD, at least in the region $\Delta x$ $\sim$ $\tau$.
This means that some model for the shrinkage-expansion mechanism has to
be introduced. In this work we use four versions of $\sigma^{str}$.
Two of them having linear and quadratic dependence of the cross section
on $\Delta x$/$\tau$,
were taken from Ref.~\cite{farrar}.
Let us briefly discuss the physical reason behind linear and quadratic
dependence (see Ref.~\cite{kopel3}). The QCD lattice calculations show
that the confinement radius is much smaller than the mean hadronic radii.
Consequently the color field in the hadrons is located in tubes with a
transverse size much smaller than the longitudinal one. The valence
quarks and diquarks are placed at the end-points of these tubes. In case
of inelastic scattering, the interacting hadron-tubes intersect in the
impact parameter plane. The probability of the crossing of the tubes is
proportional to their length. This means that $\sigma^{str}$ increases
proportional to $\Delta x$/$\tau$. In the naive parton model, the
inelastic cross section of 
a hadron with a nucleon is proportional to the transverse area which is 
filled in by its partons, i.e. $\sigma^{str}$ increases proportional 
to ($\Delta x$/$\tau$)$^2$.
 The first version of $\sigma^{str}$ is based on quantum diffusion:
\begin{eqnarray} 
\nonumber
\sigma^{str}(\Delta x) = \theta (\tau- \Delta x)[\sigma_q +(\sigma_h-\sigma_q)
\times
\end{eqnarray}
\begin{eqnarray}
\Delta x/ \tau ] + \theta (\Delta x - \tau)\sigma_h \hspace{0.2cm},
\label{eq:equ7}
\end{eqnarray}
where $\tau = \tau_c + c \Delta \tau$, $\Delta \tau = \tau_h - \tau_c$.
We introduce the parameter $c$ (0$<c<$1) in order to take into account a
well known fact, that the string starts to interact with hadronic cross
section soon after creation of the first constituent quark of the final
hadron and before creation of second constituent.\\
\hspace*{1em}
The second version follows from naive parton case:
\begin{eqnarray}
\nonumber
\sigma^{str}(\Delta x) = \theta (\tau- \Delta x)[\sigma_q +
\end{eqnarray}
\begin{eqnarray}
(\sigma_h-\sigma_q)(\Delta x/\tau)^2 ] + \theta (\Delta x - \tau)\sigma_h
\hspace{0.2cm}.
\label{eq:equ8}
\end{eqnarray}
Two other expressions for $\sigma^{str}$ were also used ~\cite{bialas1,badalyan}:
\begin{eqnarray} 
\sigma^{str}(\Delta x) = \sigma_h -
(\sigma_h-\sigma_q)exp\Big(-\frac{\Delta x}{\tau}\Big) 
\label{eq:equ9}
\end{eqnarray}
and:
\begin{eqnarray}
\sigma^{str}(\Delta x) = \sigma_h -
(\sigma_h-\sigma_q)exp\Big(-\Big(\frac{\Delta x}{\tau}\Big)^2\Big) .
\label{eq:equ10}
\end{eqnarray}
One can easily note that at $\Delta x/\tau\ll1$ the expressions (9)
and (10) turn into (7) and (8), respectively.
At the first glance it may seems that the ITSM, as opposed to the TSM, is
actually a 1-scale model. But one must note that $\tau$ is a function of
two scales $\tau = (1 - c)\tau_c + c\tau_h$ whereas the parameter $c$
regulates inclusion of each scale into $\tau$.
\section{\label{sec:fit} Details of fit and \\results}
\hspace*{1em}
For the fit the semi-inclusive data~\cite{airapet4}
of HERMES experiment on four nuclear
targets (helium, neon, krypton, xenon) and deuterium were used. Only most precise
(high statistic) part of data was used. It was consisting
from two pieces:\\
\hspace*{1em}
(i) the piece of the one dimensional data including the $\nu$ - and
$z$ - dependences of $\pi^+$ and $\pi^-$ mesons.
Each dependence consists from 9 experimental
points, i.e. for this piece we have all together 144 points.
The one dimensional data for nuclear multiplicity ratio
are a functions of single variable $\nu$ or
$z$, whereas in model, $R_{M}^{h}$ enters as a function of
two variables $\nu$ and $z$ (the usage of two variables allows one to
avoid the problem of additional integration over $z$ or $\nu$ in eq.(1)).
For this reason we introduce in the $R_{M}^{h}$,
in case of one dimensional data, second variable by next way.
In case of $\nu$ - dependence, for each measured $\nu$ bin the value of $\hat{z}$
(averaged over the given $\nu$ bin), and
in case of $z$ - dependence, for each measured $z$ bin the value of $\hat{\nu}$ 
(averaged over the given $z$ bin) are taken from the experimental data;\\
\hspace*{1em}
(ii) the piece of the two
dimensional data, containing the charged pions data on the same nuclei.
This part is available in form of
detailed binning over $\nu$ ($z$) and three slices over $z$ ($\nu$).
We would like to remind that slices over $z$ are:
first $0.2 < z < 0.4$, second $0.4 < z < 0.7$ and third
$z > 0.7$ and over $\nu$: first $6 < \nu < 12 GeV$, second
$12 < \nu < 17 GeV$ and third $17 < \nu < 23.5 GeV$.
Each slice of each dependence consists from 8 experimental
points, besides third slices over $\nu$ in $z$-dependence, which consist
from 7 experimental points,
i.e. in this piece we have 188 points.\\
\hspace*{1em}
For one and two dimensional data we select all together 332 experimental points.
\hspace*{1em}
Now let us turn to the discussion of the ingredients of the string model.
One of the important parameters is
the string tension (string constant) which determines the energy loss
by leading quark on unit length. In this work it
was fixed at a static value determined by the Regge
trajectory slope~\cite{kopel4,sjostrand}
\begin{eqnarray}
\kappa = 1/(2\pi\alpha'_R) = 1 GeV/fm \hspace{0.2cm}.
\label{eq:equ11}
\end{eqnarray}
\begin{table}[t!]
\begin{tabular}{|c|c|c|c|}
\hline
\multicolumn{4}{|c|}{$\sigma^{str}(7)$} \\
\hline
$NDF$ & $\sigma_{q} \hspace{0.2cm} mb$ & $c$ & $\hat{\chi}^2$ \\
\hline
1 & $0.62 \pm 0.15$ & $0.247 \pm 0.017$ & 0.63 \\
\hline
2 & $0.81 \pm 0.16$ & $0.231 \pm 0.019$ & 0.57 \\
\hline
3 & $0.91 \pm 0.16$ & $0.202 \pm 0.017$ & 0.61 \\
\hline
\multicolumn{4}{|c|}{$\sigma^{str}(8)$} \\
\hline
$NDF$ & $\sigma_{q} \hspace{0.2cm} mb$ & $c$ & $\hat{\chi}^2$ \\
\hline
1 & $3.71 \pm 0.10$ & $0.175 \pm 0.010$ & 0.55 \\
\hline
2 & $3.90 \pm 0.10$ & $0.161 \pm 0.009$ & 0.50 \\
\hline
3 & $4.12 \pm 0.12$ & $0.140 \pm 0.014$ & 0.57 \\
\hline
\multicolumn{4}{|c|}{$\sigma^{str}(9)$} \\
\hline
$NDF$ & $\sigma_{q} \hspace{0.2cm} mb$ & $c$ & $\hat{\chi}^2$ \\
\hline
1 & $1.29 \pm 0.12$ & $0.079 \pm 0.012$ & 0.55 \\
\hline
2 & $1.53 \pm 0.13$ & $0.065 \pm 0.012$ & 0.53 \\
\hline
3 & $1.76 \pm 0.14$ & $0.040 \pm 0.012$ & 0.59 \\
\hline
\multicolumn{4}{|c|}{$\sigma^{str}(10)$} \\
\hline
$NDF$ & $\sigma_{q} \hspace{0.2cm} mb$ & $c$ & $\hat{\chi}^2$ \\
\hline
1 & $3.91 \pm 0.10$ & $0.100 \pm 0.012$ & 0.56 \\
\hline
2 & $4.13 \pm 0.11$ & $0.087 \pm 0.012$ & 0.53 \\
\hline
3 & $4.39 \pm 0.12$ & $0.066 \pm 0.012$ & 0.60 \\
\hline
\end{tabular}
\vskip 0.5cm
\caption{\label{tab:tab1} {\it Values of fitting parameters and
$\hat{\chi}^{2}$ in case of $\tau_{c}(3)$ and total
errors. For NDF from eq.(13) versions 1, 2, 3
are sets of parameters from eqs.(14), (15), (16), respectively.
$\sigma^{str}(7)$ means $\sigma^{str}$ from eq.(7) etc.}}
\end{table}
\begin{table}[ht!]
\begin{tabular}{|c|c|c|c|}
\hline
\multicolumn{4}{|c|}{$\sigma^{str}(7)$} \\
\hline
$NDF$ & $\sigma_{q} \hspace{0.2cm} mb$ & $c$ & $\hat{\chi}^2$ \\
\hline
1 & $0.00 \pm 0.01$ & $0.372 \pm 0.012$ & 1.89 \\
\hline
2 & $0.00 \pm 0.01$ & $0.343 \pm 0.011$ & 1.71 \\
\hline
3 & $0.00 \pm 0.01$ & $0.313 \pm 0.012$ & 1.67 \\
\hline
\multicolumn{4}{|c|}{$\sigma^{str}(8)$} \\
\hline
$NDF$ & $\sigma_{q} \hspace{0.2cm} mb$ & $c$ & $\hat{\chi}^2$ \\
\hline
1 & $1.27 \pm 0.11$ & $0.146 \pm 0.004$ & 1.01 \\
\hline
2 & $1.41 \pm 0.22$ & $0.137 \pm 0.018$ & 0.91 \\
\hline
3 & $1.28 \pm 0.23$ & $0.113 \pm 0.016$ & 0.92 \\
\hline
\multicolumn{4}{|c|}{$\sigma^{str}(9)$} \\
\hline
$NDF$ & $\sigma_{q} \hspace{0.2cm} mb$ & $c$ & $\hat{\chi}^2$ \\
\hline
1 & $0.00 \pm 0.01$ & $0.111 \pm 0.009$ & 0.97 \\
\hline
2 & $0.00 \pm 0.01$ & $0.085 \pm 0.009$ & 0.85 \\
\hline
3 & $0.00 \pm 0.01$ & $0.057 \pm 0.009$ & 0.84 \\
\hline
\multicolumn{4}{|c|}{$\sigma^{str}(10)$} \\
\hline
$NDF$ & $\sigma_{q} \hspace{0.2cm} mb$ & $c$ & $\hat{\chi}^2$ \\
\hline
1 & $1.82 \pm 0.14$ & $0.077 \pm 0.010$ & 0.74 \\
\hline
2 & $1.99 \pm 0.14$ & $0.068 \pm 0.010$ & 0.67 \\
\hline
3 & $2.03 \pm 0.16$ & $0.047 \pm 0.010$ & 0.71 \\
\hline
\end{tabular}
\vskip 0.5cm
\caption{\label{tab:tab2} {\it Values of fitting parameters and
$\hat{\chi}^{2}$ in case of $\tau_{c}(4)$ and total
errors.}}
\end{table}

\begin{table}[ht!]
\begin{tabular}{|c|c|c|c|}
\hline
\multicolumn{4}{|c|}{$\sigma^{str}(7)$} \\
\hline
$NDF$ & $\sigma_{q} \hspace{0.2cm} mb$ & $c$ & $\hat{\chi}^2$ \\
\hline
1 & $0.87 \pm 0.04$ & $0.256 \pm 0.007$ & 9.34 \\
\hline
2 & $1.07 \pm 0.03$ & $0.241 \pm 0.006$ & 7.47 \\
\hline
3 & $1.19 \pm 0.04$ & $0.202 \pm 0.006$ & 9.32 \\
\hline
\multicolumn{4}{|c|}{$\sigma^{str}(8)$} \\
\hline
$NDF$ & $\sigma_{q} \hspace{0.2cm} mb$ & $c$ & $\hat{\chi}^2$ \\
\hline
1 & $3.87 \pm 0.02$ & $0.165 \pm 0.004$ & 8.49 \\
\hline
2 & $4.07 \pm 0.03$ & $0.150 \pm 0.005$ & 6.55 \\
\hline
3 & $4.32 \pm 0.03$ & $0.123 \pm 0.005$ & 9.06 \\
\hline
\multicolumn{4}{|c|}{$\sigma^{str}(9)$} \\
\hline
$NDF$ & $\sigma_{q} \hspace{0.2cm} mb$ & $c$ & $\hat{\chi}^2$ \\
\hline
1 & $1.49 \pm 0.03$ & $0.066 \pm 0.004$ & 8.31 \\
\hline
2 & $1.73 \pm 0.03$ & $0.052 \pm 0.004$ & 7.02 \\
\hline
3 & $1.99 \pm 0.03$ & $0.020 \pm 0.004$ & 9.15 \\
\hline
\multicolumn{4}{|c|}{$\sigma^{str}(10)$} \\
\hline
$NDF$ & $\sigma_{q} \hspace{0.2cm} mb$ & $c$ & $\hat{\chi}^2$ \\
\hline
1 & $4.05 \pm 0.02$ & $0.081 \pm 0.004$ & 8.91 \\
\hline
2 & $4.26 \pm 0.02$ & $0.068 \pm 0.004$ & 7.36 \\
\hline
3 & $4.56 \pm 0.03$ & $0.042 \pm 0.004$ & 9.80 \\
\hline
\end{tabular}
\vskip 0.5cm
\caption{\label{tab:tab3} {\it Values of fitting parameters and
$\hat{\chi}^{2}$ in case of $\tau_{c}(3)$ and statistical
errors.}}
\end{table}
\begin{table}[ht!]
\begin{tabular}{|c|c|c|c|}
\hline
\multicolumn{4}{|c|}{$\sigma^{str}(7)$} \\
\hline
$NDF$ & $\sigma_{q} \hspace{0.2cm} mb$ & $c$ & $\hat{\chi}^2$ \\
\hline
1 & $0.00 \pm 0.00$ & $0.526 \pm 0.005$ & 38.3 \\
\hline
2 & $0.00 \pm 0.00$ & $0.486 \pm 0.004$ & 34.8 \\
\hline
3 & $0.00 \pm 0.00$ & $0.441 \pm 0.004$ & 34.8 \\
\hline
\multicolumn{4}{|c|}{$\sigma^{str}(8)$} \\
\hline
$NDF$ & $\sigma_{q} \hspace{0.2cm} mb$ & $c$ & $\hat{\chi}^2$ \\
\hline
1 & $0.93 \pm 0.04$ & $0.141 \pm 0.004$ & 21.6 \\
\hline
2 & $1.14 \pm 0.04$ & $0.141 \pm 0.004$ & 19.6 \\
\hline
3 & $0.83 \pm 0.05$ & $0.099 \pm 0.004$ & 20.2 \\
\hline
\multicolumn{4}{|c|}{$\sigma^{str}(9)$} \\
\hline
$NDF$ & $\sigma_{q} \hspace{0.2cm} mb$ & $c$ & $\hat{\chi}^2$ \\
\hline
1 & $0.00 \pm 0.00$ & $0.157 \pm 0.003$ & 21.4 \\
\hline
2 & $0.00 \pm 0.00$ & $0.121 \pm 0.003$ & 18.2 \\
\hline
3 & $0.00 \pm 0.00$ & $0.079 \pm 0.003$ & 18.5 \\
\hline
\multicolumn{4}{|c|}{$\sigma^{str}(10)$} \\
\hline
$NDF$ & $\sigma_{q} \hspace{0.2cm} mb$ & $c$ & $\hat{\chi}^2$ \\
\hline
1 & $1.54 \pm 0.03$ & $0.059 \pm 0.003$ & 15.8 \\
\hline
2 & $1.73 \pm 0.04$ & $0.053 \pm 0.003$ & 13.9 \\
\hline
3 & $1.70 \pm 0.04$ & $0.023 \pm 0.004$ & 15.3 \\
\hline
\end{tabular}
\vskip 0.5cm
\caption{\label{tab:tab4} {\it Values of fitting parameters and
$\hat{\chi}^{2}$ in case of $\tau_{c}(4)$ and statistical
errors.}}
\end{table}
For calculations the following
nuclear density functions (NDF) were used. For deuterium
the hard core deuteron wave functions from Ref.~\cite{reid} were used. For
$^4He$ the shell
model~\cite{elton} was used:
\begin{eqnarray}
\rho(r)=\rho_0(\frac{4}{A}
+\frac{2}{3}\frac{(A-4)}{A}\frac{r^2}{r_A^2})exp(-\frac{r^2}{r_A^2}) ,
\label{eq:equ12}
\end{eqnarray}
where $r_A=1.31~fm$ for $^4He$.  
For $^{20}Ne$, $^{84}Kr$ and $^{131}Xe$ the Woods-Saxon distribution was
used
\begin{eqnarray}
\rho(r) = \rho_0 /(1+exp((r-r_A)/a)) \hspace{0.2cm}.
\label{eq:equ13}
\end{eqnarray}
Three sets of parameters for NDF's from eq.(13)were used for the fit:\\
first set~\cite{bialas3},  $a=0.54~fm$ ,
\begin{eqnarray}
r_A =(0.978 + 0.0206A^{1/3})A^{1/3}~fm \hspace{0.2cm};
\label{eq:equ14}
\end{eqnarray}
second set~\cite{bialas4}, $a=0.54~fm$ ,
\begin{eqnarray}
r_A =(1.19A^{1/3} -\frac{1.61}{A^{1/3}})~fm \hspace{0.2cm};
\label{eq:equ15}
\end{eqnarray}
third set~\cite{capella},  $a=0.545~fm$ ,
\begin{eqnarray}
r_A =1.14A^{1/3}~fm \hspace{0.2cm}\hspace{0.2cm}.
\label{eq:equ16}
\end{eqnarray}
The corresponding values of $\rho_0$ were determined from the normalization 
condition:  
\begin{eqnarray}
\int{d^3r \rho(r)} = 1 \hspace{0.2cm}.
\label{eq:equ17}
\end{eqnarray}
Parameter $a$ is practically the same for all three sets, radius
$r_A$ for the third set is larger by approximately 6~\%  than the ones for
the first and second sets.
Let us briefly discuss the choice of the nuclear matter distribution
functions. For deuterium the choice of the NDF
is not important because the FSI are small.
For light nucleus $^{4}He$ the shell model was used,
because there was no alternative.
For middle and heavy nuclei preferable NDF is Woods-Saxon distribution.
However, there is some freedom in the choice of the parameters
themselves, therefore we have included three sets of parameters (14)-(16),
in order to study uncertainty of the fitting procedure related to the NDFs.
Two expressions for $\tau_c$ were used for the
fit - equations (3) and (4). For $\sigma^{str}(\Delta x)$
four different expressions from
eqs.(7)-(10) were used. The values of $\sigma_h$ (hadron-nucleon inelastic
cross section) used in the fit were set equal to: $\sigma_{\pi^+} =   
\sigma_{\pi^-} = 20~mb$. The same value of inelastic cross section was
used for charged pions.\\
\hspace*{1em}
The fit was performed to tune two parameters: the initial value of
string-nucleon cross section $\sigma_{q}$ and coefficient $c$.\\
\hspace*{1em}

The quantitative criterium $\hat{\chi}^2$ was
used. As usually it was determined as:
\begin{eqnarray}
\nonumber
\hat{\chi}^2 = \frac{1}{(n_{exp}-n_{par}-1)}\times
\end{eqnarray}
\begin{eqnarray}
\nonumber
\sum_{n=1}^{n_{exp}}
\Big(\frac{R_{M}^{h}(theor)-R_{M}^{h}(exp)}{\Delta R_{M}^{h}(exp)} \Big)^{2} 
\hspace{0.2cm},
\end{eqnarray}
where $n_{exp}$ and $n_{par}$ are numbers of experimental points and parameters;
$R_{M}^{h}(theor)$ is the theoretical value for ratio at given point;
$R_{M}^{h}(exp)$ and $\Delta R_{M}^{h}(exp)$ are experimental values of
$R_{M}^{h}$ and its error.
Let us firstly discuss fit with use of total errors. Results are
presented in the Tables~\ref{tab:tab1} and~\ref{tab:tab2}.
Easily to see that ITSM describes data on quantitative level. The version
with $\tau_{c}(~\ref{eq:equ3})$ gives for $\hat{\chi}^2$ the values in
order of $0.5$ for all versions of $\sigma^{str}$. The version with
$\tau_{c}(~\ref{eq:equ4})$ gives for $\hat{\chi}^2$ the values in order
of unity. The common tendency is that $\sigma^{str}$ is essentially smaller
than hadron-nucleon cross sections and coefficient $c$ essentially smaller
than unity.
The obtained values of $\hat{\chi}^2$ show, that total errors are still
large and do
not allow to verify the different versions of ITSM. It is a reason why we
turn to the fit of the data with statistical errors only.
Corresponding results are presented in the Tables~\ref{tab:tab3}
and~\ref{tab:tab4}. Again, as in case of total errors the values of
$\hat{\chi}^2$ for $\tau_{c}(~\ref{eq:equ3})$ approximately two times
smaller than for $\tau_{c}(~\ref{eq:equ4})$ but now values of
$\hat{\chi}^2$ larger as minimum ten times in comparison with preceding
case. For version with $\tau_{c}$ in form of eq.(3) minimum
$\hat{\chi}^2 = 6.55$ is obtained in case of $\sigma^{str}$ corresponding
to eq.(8) and NDF in form of eq.(15). For version with $\tau_{c}$ in
form of eq.(4)
minimum $\hat{\chi}^2 = 13.9$ is obtained in case of $\sigma^{str}$
corresponding to eq.(10) and NDF in form of eq.(15). At the first glance
$\hat{\chi}^2$
are too large and it is impossible to speak about quantitative description
of data. But we would like to remember that statistical errors of data obtained by
HERMES experiment and using for this fit~\cite{airapet4} are in order of 1~\%
and for some points even 0.5~\%. For experimental points
measured with such small errors we obtain $\hat{\chi}^2 = 9$ when difference
between theoretical model and experimental value is 1.5~\% only.
We think, that $\hat{\chi}^2 = 6.55$ which is obtained in case of
$\tau_{c}(3)$, $\sigma^{str}(8)$ and NDF(15) is suitable for
description of experimental data. The agreement of version with $\tau_{c}(4)$
is worse. We will discuss comparison of these two versions in the next section.\\
\hspace*{1em}
Except the basic fit
which was discussed above, three another fits were also performed. First of them
was performed with the same data set as basic one but three
parameters were used. Two of them were taken the same as in basic fit and
as a third parameter the string tension $\kappa$ was used. The results of fit with
$\tau_{c}(~\ref{eq:equ3})$ were obtained very close to the basic fit
for both fitting parameters and $\hat{\chi}^2$.
Indeed, for $\sigma^{str}(8)$ and NDF in form of eq.(15) the values
$\sigma_{q} = 3.87 \pm 0.042 mb$, $c = 0.18 \pm 0.01$,
$\kappa = 1.075 \pm 0.013 GeV/fm$ and 
$\hat{\chi}^2 = 6.54$ were obtained.
In this case inclusion of third parameter does not felt. The results of fit with
$\tau_{c}(~\ref{eq:equ4})$ show the stronger dependence from the third parameter.
For this case the following values for $\sigma^{str}(10)$ and
NDF(15) were obtained:\\
$\sigma_{q} = 3.36 \pm 0.058 mb$, $c = 0.00 \pm 0.00$,
$\kappa = 0.73 \pm 0.008 GeV/fm$ and $\hat{\chi}^2 = 11.7$. 
It is easily to see that the
values of parameters significantly differ from the corresponding values in basic
fit, but $\hat{\chi}^2$ is close enough to the one for basic fit.
Two other fits were performed
separately for one and two dimensional pieces of data. Results
are close to the ones for basic fit. When data are
divided on one and two dimensional parts, the $\hat{\chi}^2$
for one dimensional data are obtained slightly larger than for two dimensional
part because in first case statistical errors are smaller. But general
situation is very resemble to the one for basic fit.\\
\hspace*{1em}
Let us compare the procedure and results of present fit with the
our preceding fit~\cite{akopov2}. At that time the SIDIS data of
HERMES experiment
for two targets: nitrogen~\cite{airapet1} and krypton~\cite{airapet2} were available.
The $\nu$- and $z$-dependences of
$\pi^{+}$ and $\pi^{-}$ mesons with 
58 available experimental points were included in the fit procedure. As the experimental errors only statistical
ones were 
taken. Minimum values of $\hat{\chi}^2$ (best fit)
for ITSM two parameter's fit were obtained for
$\tau_{c}$ from eq.(3) ($\hat{\chi}^2 = 1.4$) and for $\tau_{c}$ from eq.(4)
($\hat{\chi}^2 = 1.5$). Comparison with the new fit showes that $\hat{\chi}^2$
became several times larger. It should be pointed out that reasons for this increase
are: (i) very small values of statistical errors of present data (as a
minimum two times smaller than of preceding data); (ii) in the preceding
fit were included data for two nuclei only (nitrogen and krypton). The data
on nitrogen had smaller number of experimental points and
essentially larger errors than data on krypton, i.e. fit was mainly based on krypton
data. In present case data are available for four nuclei from light to heavy
(helium, neon, krypton and xenon)
and these data have comparable statistics.  
\section{\label{sec:res} Comparison with data and discussion}
\hspace*{1em}
The results of the performed fit are presented in Tables~\ref{tab:tab1},
~\ref{tab:tab2},~\ref{tab:tab3} and~\ref{tab:tab4}.
\begin{figure}[!htb]
\begin{center}
\epsfxsize=7.cm
\epsfbox{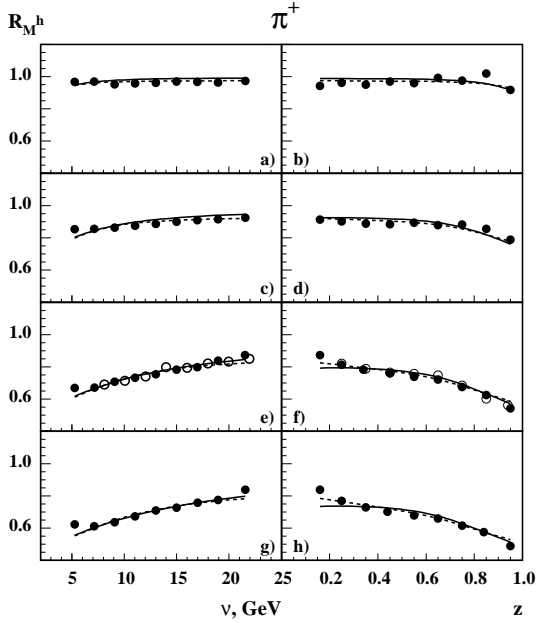}
\end{center}
\caption{\label{fig:fig2} {\it One dimensional data. Hadron multiplicity
ratio $R_{M}^{h}$ for $\pi^{+}$ mesons as a function of variable $\nu$
(left panels) and $z$ (right panels). The results are presented for
$^{4}He$ (a, b), $^{20}Ne$ (c, d), $^{84}Kr$ (e, f) and $^{131}Xe$ (g, h).
Experimental points from Ref.~\cite{airapet4} (filled circles) and
Ref.~\cite{airapet2} (open circles). The curves were
calculated with the best values of parameters obtained for the constituent
formation length in form of eq.(3) (dashed curves) and eq.(4) (solid curves).
These data were included in fit.}}
\end{figure}
\begin{figure}[!htb]
\begin{center}
\epsfxsize=7.cm
\epsfbox{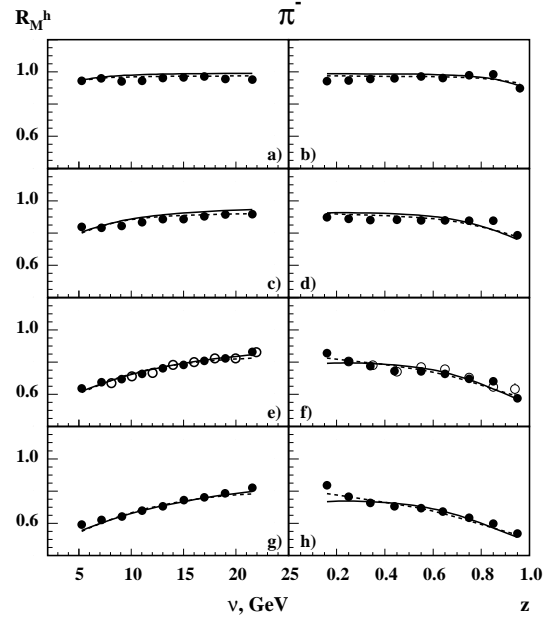}
\end{center}
\caption{\label{fig:fig3} {\it
The same as described in the caption of the
Fig.2 done for $\pi^{-}$ mesons. These data were included in fit.}}   
\end{figure}
Tables~\ref{tab:tab1} and~\ref{tab:tab2} show the best values of the fitted
parameters, their errors and $\hat{\chi}^2$ for the case of total errors.
Tables~\ref{tab:tab3} and~\ref{tab:tab4} show these values for the case when only
statistical errors were used for the fit procedure. Further we will discuss only 
last case.
\begin{figure}[h!]
\begin{center} 
\epsfxsize=7.cm
\epsfbox{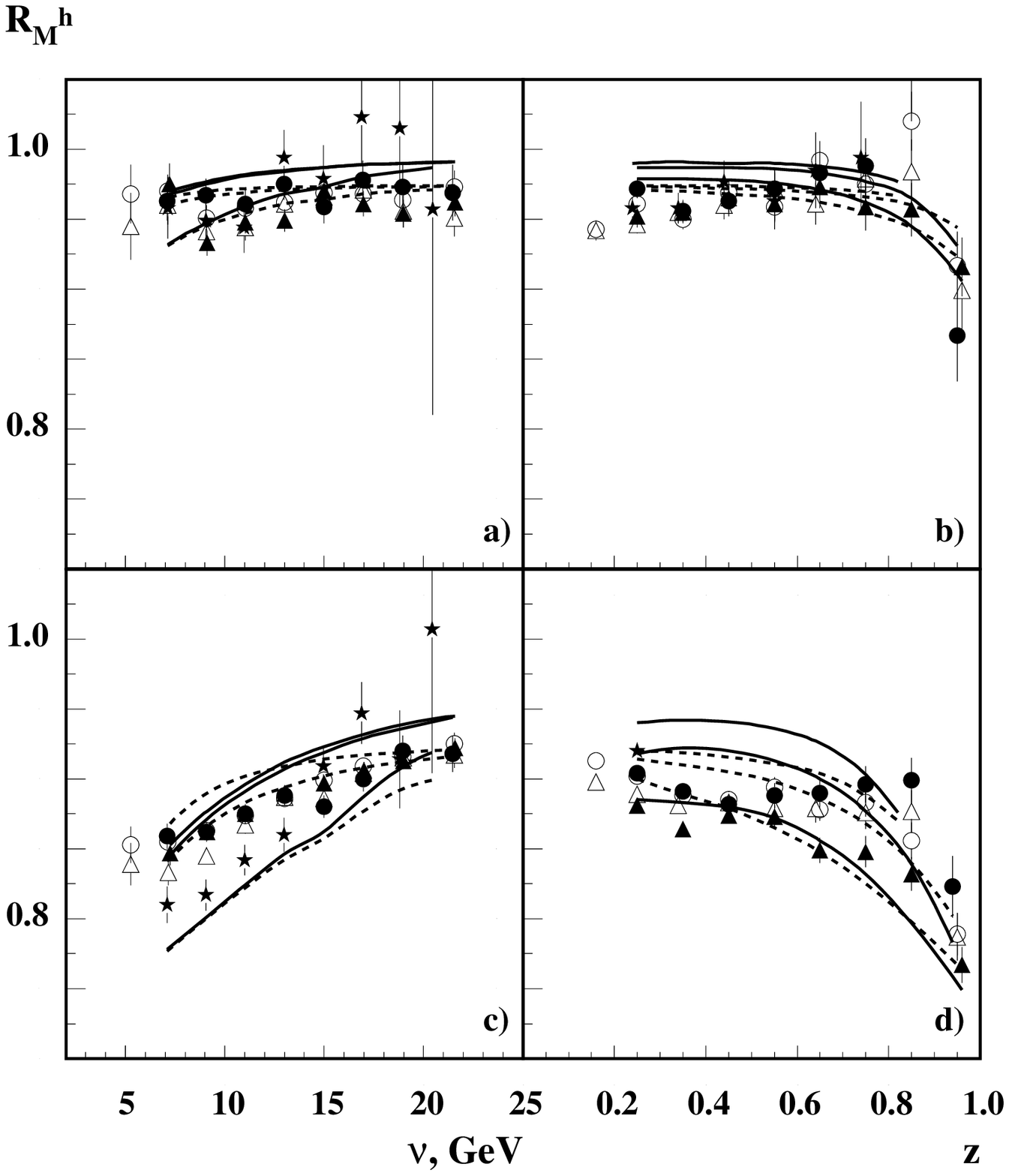}
\end{center}
\caption{\label{fig:fig4} {\it The two dimensional data.
The ratio $R_{M}^{h}$ for charged pions on $^{4}He$ (panels a, b) and
$^{20}Ne$ (c, d) nuclei as a function of
$\nu$ (left panels) and $z$ (right panels).
Experimental points from Ref.~\cite{airapet4}.
Filled symbols are: triangles - for the first slice; circles - for second slice
and stars - for third slice.
Open symbols represent one dimensional data: circles - for $\pi^{+}$ and
triangles - for $\pi^{-}$ mesons.
The theoretical curves were
calculated with the best values of parameters obtained for the constituent
formation length
in form of eq.(3) (dashed curves) and eq.(4) (solid curves).
These data were included in fit.}}
\end{figure}  
 \begin{figure}[ht!]
\begin{center}
\epsfxsize=7.cm
\epsfbox{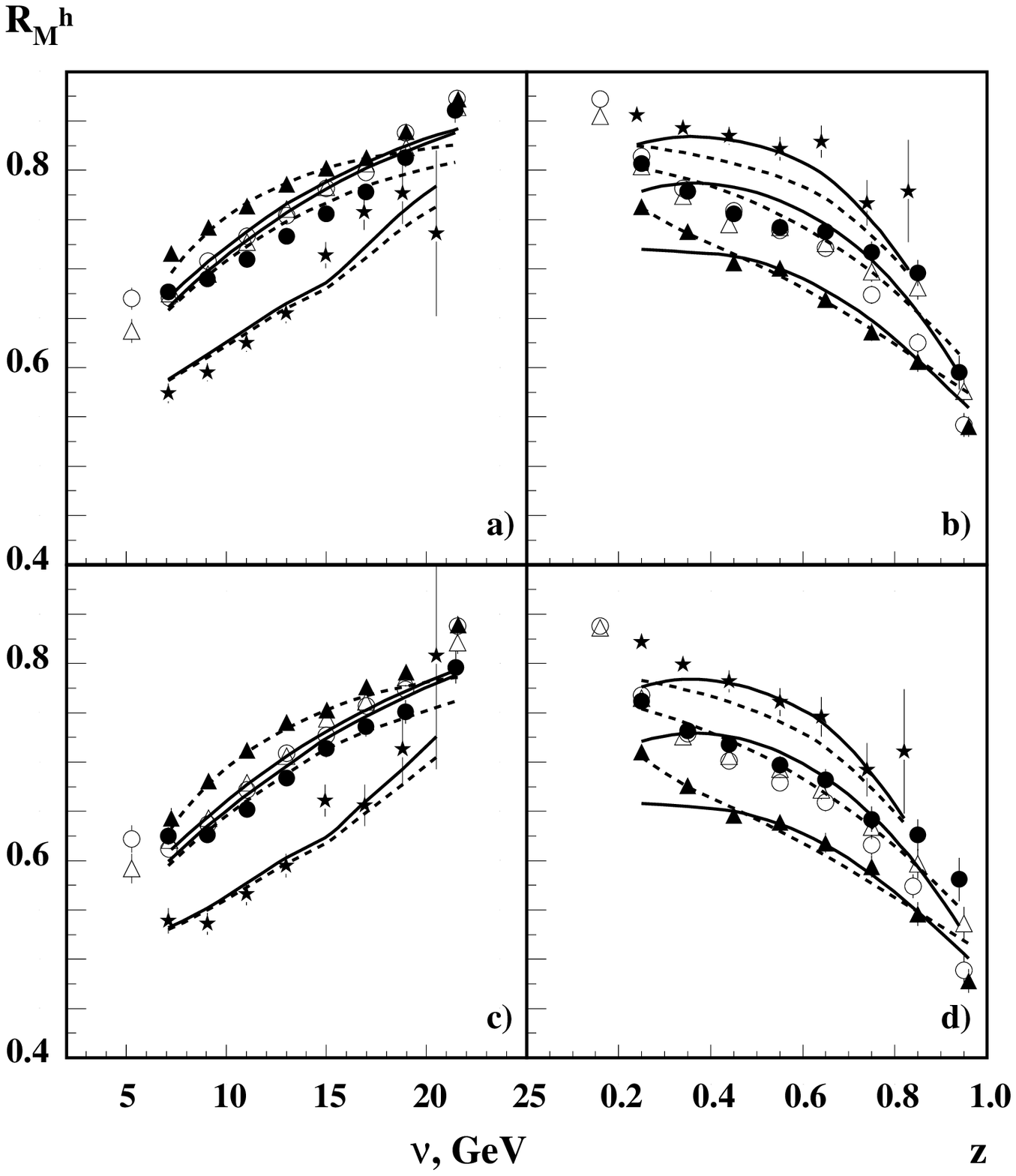}
\end{center}
\caption{\label{fig:fig5}{\it The same as described in the caption of the
Fig.~\ref{fig:fig4} done for $^{84}Kr$ (panels a, b) and $^{131}Xe$ (c,d)
targets. These data were included
in fit.}}
\end{figure}
In Figs.~\ref{fig:fig2} and~~\ref{fig:fig3} the one dimensional data
for the hadron multiplicity ratio $R_{M}^{h}$ for $\pi^{+}$ and
$\pi^{-}$ mesons as a function of variable $\nu$ (left panels) and $z$   
(right panels) are presented. The results for $^{4}He$(panels a and b),
$^{20}Ne$ (c, d), $^{84}Kr$ (e, f) and $^{131}Xe$ (g, h) are presented.
Experimental points were taken from Ref.~\cite{airapet4} (filled circles).
For comparison we presented in case of krypton previous HERMES
data~\cite{airapet2} (open circles) also.
The theoretical curves
were calculated with the best values of parameters obtained for the
versions of model with constituent formation length in form of eq.(3)
(dashed curves) and eq.(4) (solid curves). Version with
$\tau_{c}$ in form of eq.(3) describes data
presented in Figs.~\ref{fig:fig2} and~~\ref{fig:fig3} better
than version with $\tau_{c}$ in form of eq.(4), but size of presented figures
does not allow to see this difference.\\
\hspace*{1em}
In Fig.~\ref{fig:fig4} the two dimensional data for
$R_{M}^{h}$ for charged pions on $^{4}He$ (panels a, b) and $^{20}Ne$ (panels c, d)
as a functions of $\nu$ (left panels) and $z$ (right panels) are presented.
In Fig.~\ref{fig:fig5} the two dimensional data for                  
$R_{M}^{h}$ for charged pions on $^{84}Kr$ (panels a, b) and $^{131}Xe$ (panels c, d)
as a functions of $\nu$ (left panels) and $z$ (right panels) are presented.
Experimental points are taken from Ref.~\cite{airapet4}.
Filled symbols are: triangles - experimental points for the first slice
over $z$ ($\nu$) for $\nu$ - ($z$ - ) dependence; circles - for the second
slice and stars - for the third slice.
The open symbols represent the one dimensional data and were included
in figures for comparison with two dimensional ones. The open circles were
chosen for $\pi^{+}$ and open triangles for $\pi^{-}$ mesons.
Unfortunately the two dimensional data for helium are not informative
because the points from different slices
are mixed. Data for $^{20}Ne$ are more useful because
the separation of points from different slices partly takes place.
The complete
separation of data from different slices takes place for heavy nuclei
(krypton and xenon). They carry additional information in comparison
with one dimensional data and are important for the development of theoretical
models. From these figures we see, that one dimensional data (open points)
mainly coincide with second slices and do not reflect the behavior of data
in first and third ones.
The theoretical curves were
calculated with the best values of parameters obtained for the constituent
formation length
in form of eq.(3) (dashed curves) and eq.(4) (solid curves).
Let us remind that $\tau_{c}$ of eq.(3) was obtained for leading hadron
while $\tau_{c}$ of eq.(4) is the average value of this quantity in the
standard Lund model.
As we already mentioned above the two dimensional data for helium presented
in Fig.~\ref{fig:fig4}
are less useful for comparison of different versions of model. We can
state only, that theoretical curves do not contradict data.
For middle nucleus (neon) presented in Fig.~\ref{fig:fig4}, despite on
partial mixing of experimental points from different slices, we can state,
that version with $\tau_{c}$ corresponding to eq.(3) describes data better than 
one with $\tau_{c}$ corresponding to eq.(4).
And, at last, let us turn to the heavy nuclei (krypton and xenon) where
situation is more clear.
From Fig.~\ref{fig:fig5} we see that in average
the version with $\tau_{c}$ for leading hadron describes two 
dimensional data
better than one from standard Lund model. The leading hadron approach satisfactory
describes all three slices over $z$ for $\nu$ - dependence (excepting last
points), 
and also the first and second slices over $\nu$
for $z$ - dependence but underestimates data in third slice,
although the behavior is true. The main problem of the version of
model with $\tau_{c}$ taken from standard 
Lund model (eq.(4))
is the unsatisfactory description of first and second slices over $z$
in $\nu$ - dependence. This version of model practically does not differ
data with small and middle $z$. The $z$ -dependence it describes on
satisfactory level excepting the region of small $z$. It is worth to
mention that the difficulties in the description of
$\nu$ - and $z$ - dependences are mutually connected.\\
\begin{figure}[ht!]
\begin{center}
\epsfxsize=7.cm
\epsfbox{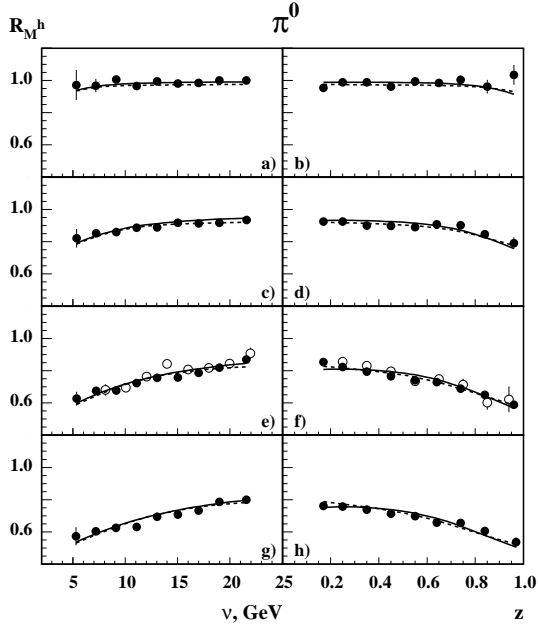}
\end{center}
\caption{\label{fig:fig6} {\it One dimensional data. Hadron multiplicity
ratio $R_{M}^{h}$ for $\pi^{0}$ mesons as a function of variable $\nu$   
(left panels) and $z$ (right panels). The results are presented for
$^{4}He$ (a, b), $^{20}Ne$ (c, d), $^{84}Kr$ (e, f) and $^{131}Xe$ (g, h).
Experimental points were taken
from Ref.~\cite{airapet4} (filled circles) and
Ref.~\cite{airapet2} (open circles). The curves were
calculated with the best values of parameters obtained for the constituent
formation length in form of eq.(3) (dashed curves) and eq.(4) (solid curves).
These data not included in fit.}}
\end{figure}
\begin{figure}[ht!]
\begin{center}
\epsfxsize=7.cm
\epsfbox{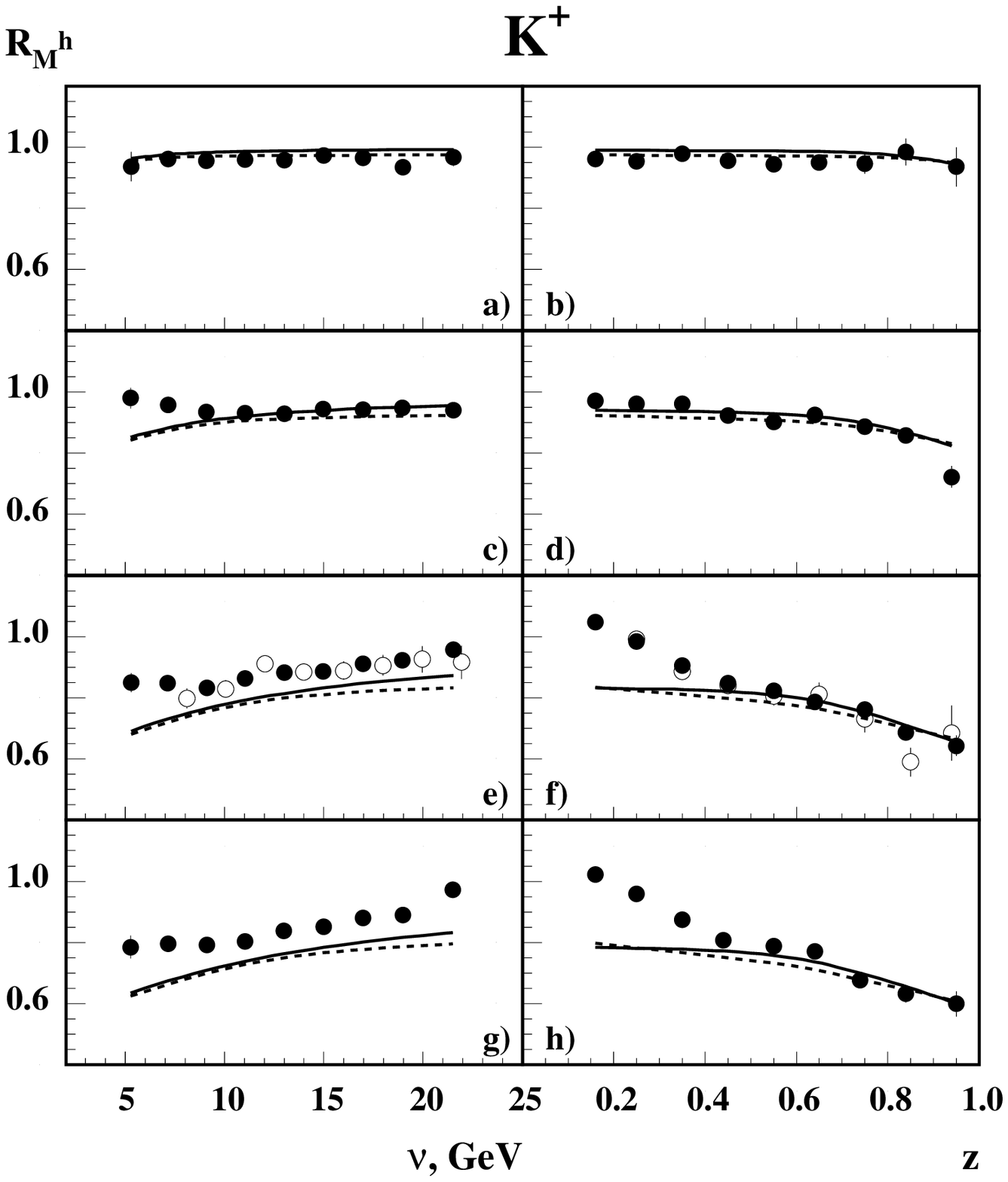}
\end{center}
\caption{\label{fig:fig7} {\it The same as described in the caption of
the Fig.~\ref{fig:fig6} done for $K^{+}$ mesons.}}
\end{figure}
\begin{figure}[ht!]
\begin{center}
\epsfxsize=7.cm
\epsfbox{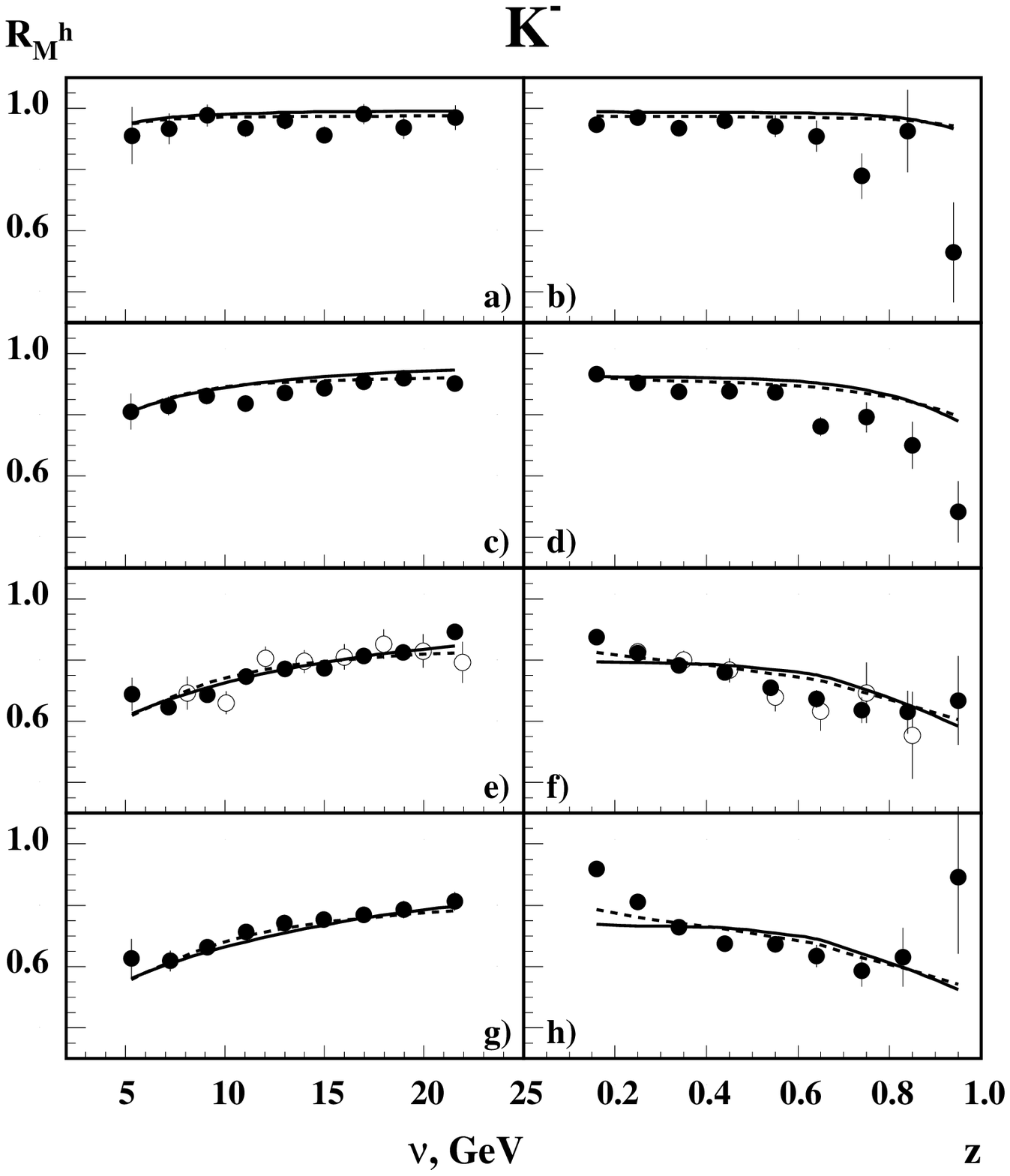}
\end{center}
\caption{\label{fig:fig8} {\it The same as described in the caption of
the Fig.~\ref{fig:fig6} done for $K^{-}$ mesons.}}
\end{figure}
\begin{figure}[ht!]
\begin{center}
\epsfxsize=7.cm
\epsfbox{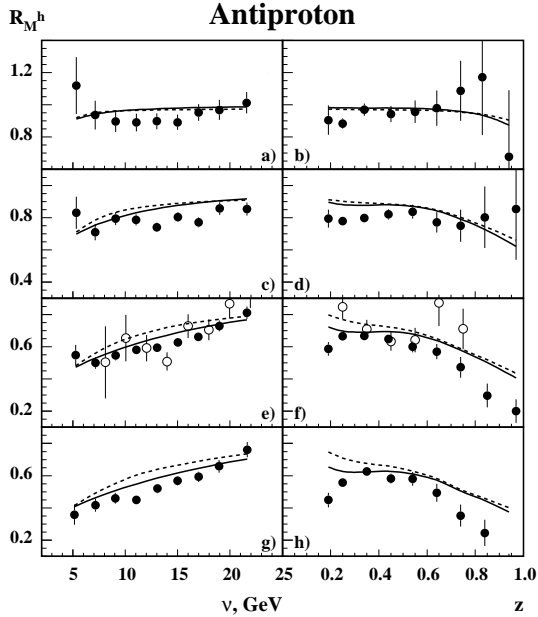}
\end{center}
\caption{\label{fig:fig9} {\it The same as described in the caption of
the Fig.~\ref{fig:fig6} done for antiproton.}}
\end{figure}
\hspace*{1em}
Furthermore, the NA for one dimensional data of
hadrons produced on all nuclear targets (but not included in fit), were
calculated in our model. In Figures 6 - 9 the $\nu$ - and
$z$ - dependences are presented for $\pi^{0}$, $K^{+}$, $K^{-}$ mesons and antiprotons,
respectively. The hadron multiplicity 
ratio $R_{M}^{h}$ for mentioned hadrons as a function of variable $\nu$
(left panels) and $z$ (right panels) are presented for 
$^{4}He$ (a, b), $^{20}Ne$ (c, d), $^{84}Kr$ (e, f) and $^{131}Xe$ (g, h).
Experimental points were taken from Ref.~\cite{airapet4} (filled circles).
For $^{84}Kr$ were taken also data from Ref.~\cite{airapet2} (open circles).
The theoretical curves were calculated with the best values of parameters
obtained for the versions with $\tau_{c}$
in form of eq.(3) (dashed curves) and eq.(4) (solid curves).
The following values of inelastic cross sections
$\sigma_h$ were used: $\sigma_{\pi^0} = \sigma_{K^-}
= 20~mb$, $\sigma_{K^+} = 14~mb$ and $\sigma_{\bar{p}} = 42~mb$.
Calculations, which were performed without additional fit, satisfactory
describe data for $\pi^{0}$ (Fig.~\ref{fig:fig6}) and
$K^{-}$ (Fig.~\ref{fig:fig8})
mesons. Here we also have, as in case of $\pi^{+}$ and $\pi^{-}$
mesons, that version with $\tau_{c}$ in form of eq.(3)
describes data slightly better than with $\tau_{c}$ in form of eq.(4).\\
\hspace*{1em}
It is worth to discuss data for $K^{+}$ mesons from Fig.~\ref{fig:fig7}
more in details, because here we have some problems. Let us begin with
$\nu$ - dependence. The data on $^{4}He$ are described very well. For $^{20}Ne$
nucleus we have disagreement between experimental data and theoretical
model for the first two points. And, at last, for heavy nuclei we have
essential underestimation of data by theoretical model. 
In this case comparison with the previous data for krypton helps to understand
that problem rather in model than in data. In case of $z$ - dependence we
have satisfactory agreement for $^{4}He$ and $^{20}Ne$.
Again description is worse for heavy nuclei. First three points are
underestimated by model. We see that in region of small $z$
experimental values of $R_{M}^{h}$ decrease, while theoretical
ones are rather constant. There is some disagreement between experimental
data and theoretical model in case of antiprotons also
(see Fig.~\ref{fig:fig9}).
In this case theoretical curves slightly overestimate data. But we do not
discuss it here in detail because of two reasons: (i) large errors and
(ii) some differences of the antiproton production mechanism from
mesons one, which do not taken into account in present paper.
\section{\label{sec:conc} Conclusions.}
The recent HERMES data~\cite{airapet4}
were used to perform the fit for ITSM.
Two-parameter's fit demonstrates satisfactory agreement with data.
The main goal of this paper was the further development of the model,
in particular the choice of proper version for $\tau_c$.
Minimum $\hat{\chi}^2$ (best fit) was obtained for version of model
with $\tau_c$ in form of eq.(3). The version with $\tau_c$ in form of
eq.(4) gave essentially larger value for $\hat{\chi}^2$.
Comparison with the two dimensional data
obtained, for the first time, by HERMES experiment, allowed to 
perform additional verification of
the different choices of constituent formation length. In particular
it was obtained, that version with $\tau_c$ in form of eq.(3) has
difficulty in description of third slice over $\nu$ in $z$ -dependence,
i.e. in region of large $\nu$. The difficulties of version
with $\tau_c$ in form of eq.(4) are more serious, because this version
does not differ small and middle $z$ in $\nu$ - dependence.
Although version of model with $\tau_c$ in form of eq.(3) describes
data on the satisfactory level, the two dimensional data show, that
we have some problem connected with the choice of the more adequate
form for constituent formation length.
More detail two dimensional data for identified hadrons
that is expected from HERMES will
provide essentially better conditions for the choice of preferable
version of the model in terms of different expressions for
$\tau_c$ and, may be, $\sigma^{str}$.
In all versions we have obtained that $\sigma_q \ll \sigma_h$.
This indicates that at early stage of hadronization process the color 
transparency takes place. It is worth to mention that the same result
was obtained in our preceding fit~\cite{akopov2}.
We do not include in consideration the NA of protons, because in this
case additional mechanisms connected with color interaction (string-flip)
and final hadron re-scattering become essential (see for instance
Refs.~\cite{gyulassy,czyzew}).

\end{document}